\documentclass[12pt]{article}
\pdfoutput=1
\usepackage{latexsym, graphicx,cite,mathrsfs}
\usepackage{amsmath}
\usepackage{amssymb}
\usepackage{amsthm}
\usepackage{subcaption,tabularx}

\usepackage[top = 1. in,bottom = 1 in, left = 1 in, right=1 in]{geometry}
\usepackage[colorlinks=true, linkcolor=blue, bookmarks=true]{hyperref}
\newcommand{\arXiv}[1]{\href{http://www.arXiv.org/abs/#1}{arXiv:#1}}
\newcommand{\doi}[2]{\href{http://dx.doi.org/#2}{#1}}

\makeatletter
\renewcommand\section{\@startsection {section}{1}{\z@}%
                  {-3.5ex \@plus -1ex \@minus -.2ex}
                  {2.3ex \@plus.2ex}%
                  {\normalfont\large\bfseries}}
\renewcommand\subsection{\@startsection{subsection}{2}{\z@}%
                   {-3.25ex\@plus -1ex \@minus -.2ex}%
                   {1.5ex \@plus .2ex}%
                   {\normalfont\bfseries}}
\makeatother

\newcommand{\beq}{\begin{equation}}
\newcommand{\eeq}{\end{equation}}
\newcommand{\ber}{\begin{array}}
\newcommand{\eer}{\end{array}}
\newcommand{\del}{\partial}

\newcommand{\al}{\alpha}
\newcommand{\be}{\beta}
\newcommand{\ph}{\varphi}
\newcommand{\bea}{\begin{eqnarray}}
\newcommand{\eea}{\end{eqnarray}}

\newcommand{\nn}{\nonumber}
\newcommand{\tsum}{{\textstyle\sum_J}}
\newcommand{\sumc}{\tsum c_J}
\newcommand{\s}{\sigma}

\begin{document}
\begin{titlepage}
\begin{flushright}
\phantom{arXiv:yymm.nnnn}
\end{flushright}
\vspace{5mm}
\begin{center}
{\LARGE\bf A set of master variables\vspace{5mm}\\ for the two-star random graph}
\vskip 15mm
{\large Pawat Akara-pipattana$^{a}$ and Oleg Evnin$^{b,c}$}
\vskip 10mm
{\em $^a$ Universit\'e Paris-Saclay, CNRS, LPTMS, 91405 Orsay, France}
\vskip 3mm
{\em $^b$ High Energy Physics Research Unit, Department of Physics, Faculty of Science, Chulalongkorn University,
10330 Bangkok, Thailand}
\vskip 3mm
{\em $^c$ Theoretische Natuurkunde, Vrije Universiteit Brussel and\\
 International Solvay Institutes, 1050 Brussels, Belgium}
\vskip 7mm
{\small\noindent {\tt pawat.akarapipattana@universite-paris-saclay.fr, oleg.evnin@gmail.com}}
\vskip 20mm
{\bf ABSTRACT}\vspace{3mm}
\end{center}

The two-star random graph is the simplest exponential random graph model with nontrivial interactions between the graph edges.
We propose a set of auxiliary variables that control the thermodynamic limit where the number of vertices $N$ tends to infinity.
Such `master variables' are usually highly desirable in treatments of `large $N$' statistical field theory problems.
For the dense regime when a finite fraction of all possible edges are filled, this construction recovers the mean-field solution of Park and Newman,
but with an explicit control over the $1/N$ corrections. We use this advantage to compute the first subleading correction to the Park-Newman result, which encodes the finite, nonextensive contribution to the free energy. For the sparse regime with a finite mean degree, we obtain a very compact derivation of the Annibale-Courtney solution, originally developed with the use of functional integrals, which is comfortably bypassed in our treatment. 

\vfill

\end{titlepage}

\section{Introduction}

Many quantum and statistical systems simplify when the number of degrees of freedom becomes large, for example the number of internal degrees of freedom of fields at each spacetime point, or the size of matrices, etc. One is often talking about `large $N$' limits \cite{largeN}, where $N$ parametrizes the number of degrees of freedom. A `holy grail' problem for such large $N$ limits is the search for `master variables' or `master fields', such that, when the original problem
is reformulated through these variables, $N$ becomes a numerical parameter in the action (or in the statistical probability distribution), and changing $N$ no longer affects the set of the degrees of freedom involved. 
As this parameter tends to infinity, the large $N$ limit is often controlled by an explicit saddle point in terms of the master variables, which condense, as a result,
to definite values.

The master variables can be of a different nature than the original variables. In the simplest case of the $O(N)$ vector model whose fundamental variables $\phi_i$ are vectors under $O(N)$ rotations, scalar fields constructed from these vectors condense to definite values at large $N$ and play the role of master variables \cite{largeN}. Similar `scalar condensation' via different applications of the Hubbard-Stratonovich transformation occurs in multi-matrix models with a large number of matrices transformed into one another by rotations \cite{larged1,larged2,larged3}, or in computations of random tensor eigenvalue distributions in a large number of dimensions \cite{tensor1,tensor2,tensor3}. The situation can be much more exotic, however. For example, for large random matrices of size $N\times N$, the resolvents, which are functions of an external spectral parameter, condense to definite values \cite{Staudacher, Eynard}. Thus, while the original variables are numbers, the master variables are functions. This is especially visible in the corresponding treatments of sparse random matrix ensembles \cite{BR,RB,MF,FM,laplace,degseq}:
starting with random matrices, one arrives at explicit functional integrals in terms of the functional master variables that acquire a saddle point structure at large $N$. Finally, in quantum gauge theories, loop variables make a prominent appearance at large $N$ and are believed to condense to definite values \cite{Polyakov,Das,Makeenko,loopnew}. Those are functionals on the loop space, while the original variables are fields in the physical spacetime.

Motivated by this diversity of large $N$ limits and the master variables that control them, we would like to revisit here the two-star random graph model.
This is probably the simplest statistical graph model with nontrivial edge interactions and nontrivial thermodynamics, formulated as a Gibbs-like maximal entropy ensemble
with controlled average number of edges and `2-stars' (which one can think of as the number of paths of length 2). By contrast, if one only controls the number of edges, one ends up with the edges being filled or not filled randomly and independently, which is known as the Erd\H{o}s-R\'enyi random graph. This model is of fundamental importance in the physics of networks \cite{newman}, providing a point of departure for more sophisticated constructions, but its thermodynamics is trivial (this is for the case of distinguishable vertices; thermodynamics of unlabeled graphs with indistinguishable vertices is nontrivial even within the Gibbs ensemble that only controls the number of edges \cite{unlabeled,ineqph}).

We thus turn to the two-star random graph on $N$ vertices and look for a set of master variables that control the large $N$ limit.
One generally expects that the master variables are invariants of the symmetries of the theory, and here we must keep in mind that, unlike $O(N)$ models,
the matrix models defining random graph ensembles are typically invariant under vertex permutations, but not under any form of continuous rotations,
leaving a much bigger set of invariants.

In the dense regime (a finite fraction of all possible edges occupied), a mean-field solution of the two-star model was developed by Park and Newman \cite{PN}.
This solution is simple, elegant and manifestly correct given the comparisons with numerics, but it does not bring under explicit control the $1/N$ corrections,
which is where the master variables proposed here lead to improvement. In the sparse regime (a finite average degree), the model was solved using auxiliary fields
by Annibale and Courtney \cite{AC}. This solution, however, goes through functional integrals at the intermediate stages. The master variables proposed here will recover it in a more compact and elementary fashion. (We mention for additional
perspective the mathematical works \cite{math1,math2} dealing with related topics, as well as some further relevant physics literature on the two-star model and its generalizations \cite{depinning,finitesize,mfanalog}.)

Our treatment will proceed with defining the relevant master variables in section~\ref{secmaster}, showing how they can be used to reproduce the Park-Newman mean-field
solution in the dense regime in section~\ref{secPN}, spelling out the computation of $1/N$ corrections to this solution in section~\ref{sec1N}, followed by a very
compact derivation of the large $N$ solution in the sparse case in section~\ref{secsparse}. We will give a physical interpretation to the condensation of master variables in this solution in section~\ref{seccond}, and then conclude with a brief summary and discussion.


\section{The two-star random graph and its auxiliary field representation}\label{secmaster}

For analytic considerations, one starts by representing graphs as adjacency matrices.
For each graph on $N$ vertices, define a real symmetric zero-diagonal $N\times N$ adjacency matrix $A_{ij}$ whose entries equal 1 if there and edge between vertices $i$
and $j$, and 0 otherwise. The probability of each graph in the two-star model is then given by
\beq\label{2stardef}
P[A]=e^{-H[A]}/Z,\qquad Z\equiv\sum_A e^{-H[A]},\qquad H[A]\equiv \al\sum_{kl}{A_{kl}}+\be\sum_{klm} A_{kl}A_{lm},
\eeq 
where $\sum_A$ implies summing over all possible adjacency matrices, that is, over all graph configurations.
This is a typical Gibbs-like maximal entropy ensemble with two `fugacities' $\al$ and $\be$. Introducing the vertex degrees $k_j\equiv \sum_l A_{jl}$, one can understand $\sum_{jl}{A_{jl}}=\sum_j k_j$ as twice the number of edges, while $\sum_{ljm} A_{lj}A_{jm}=\sum_j k_j^2$. These are the two quantities whose
expectation values are controlled by their thermodynamic conjugates $\al$ and $\be$. By linear redefinitions of $\al$ and $\be$, one can equivalently control instead the number of edges and the number of `two-star' motifs (ordered triplets of distinct nodes with at least two edges connecting them), given by $\sum_j k_j(k_j-1)$. While all of these definitions are physically equivalent, the conventions explicitly stated in (\ref{2stardef}) prove convenient for our subsequent analytic work.

To solve the ensemble (\ref{2stardef}) in the thermodynamic limit $N\to\infty$, one needs to evaluate the partition function $Z$.
Direct summation over $A$ is impossible due to the quadratic nonlinearity in $H$. We start with a simple Hubbard-Stratonovich transform, as in \cite{PN}:
\beq\label{HSAphi}
e^{-\be \sum_{km}A_{kl}A_{lm}}=\frac{1}{\sqrt{\pi}}\int_{-\infty}^\infty d\phi \,e^{-\phi^2 +2i\phi \sqrt{\be}\sum_k A_{kl}}.
\eeq
We need to introduce one such variable $\phi$ for each value of $l=1..N$, obtaining 
\beq\label{ZAphi}
Z= \frac1{\pi^{N/2}}\sum_{A}\int_{-\infty}^\infty d\phi_1\ldots d\phi_N\,e^{-\sum_l\phi_l^2}e^{\sum_{k<l} A_{kl}[-2\al+2i(\phi_k+\phi_l). \sqrt{\be}]},
\eeq
where we expressed the exponent through the independent entries $A_{kl}$ with $k<l$.
The summation over $A_{kl}$ with $k<l$ then trivializes since the summand is factorized over the entries of the adjacency matrix. One thereby
arrives at the following {\it vector model} in terms of $\phi_l$:
\beq\label{Zphi}
Z=\frac1{\pi^{N/2}}\int_{-\infty}^\infty d\phi_1\ldots d\phi_N\,e^{-\sum_k\phi_k^2}\prod_{k<l}\left(1+e^{-2\al+2i(\phi_k+\phi_l) \sqrt{\be}}\right),
\eeq
which is often more convenient to represent as
\beq\label{vecmoddef}
Z=\frac1{\pi^{N/2}}\int_{-\infty}^\infty d\phi_1\ldots d\phi_N\,e^{-\sum_k\phi_k^2\,+\,S[\phi]},\qquad S[\phi]\equiv\sum_{k<l}\log\left(1+e^{-2\al+2i(\phi_k+\phi_l) \sqrt{\be}}\right).
\eeq

One expects that such vector models simplify when $N$ is large. This is what in fact happens, though the details depend crucially on the $N$-scalings of the thermodynamic parameters $\al$ and $\be$ that determine whether the graph is in the sparse or dense regime (both of which are an option at large $N$, defining thus different large $N$ limits of the model). 

A mean-field solution of (\ref{vecmoddef}) in the dense regime was originally presented in \cite{PN}. The dense regime corresponds to $\al=O(1)$, $\be=O(1/N)$ at $N\to\infty$. In this regime, $\phi_k$ condense to definite values of order $\sqrt{N}$ (in the current conventions) plus fluctuations of order 1. The idea of \cite{PN} is then to expand around this condensation point as $\phi_k=\sqrt{N}\phi_0+\ph_k$ and use the Taylor expansion for $S$:
\beq\label{phiTay}
-\sum_k\phi_k^2+S[\phi]=S_0+\sum_k \phi_k^2 +\frac12\sum_{kl}\frac{\del^2 S}{\del\phi_k\del\phi_l}\ph_k\ph_l+\frac16\sum_{klm}\frac{\del^3 S}{\del\phi_k\del\phi_l\del\phi_m}\ph_l\ph_l\ph_m+\cdots.
\eeq
The mean-field value $\phi_0$ is chosen to ensure that the term linear in $\ph$ is absent. As an estimate for $Z$ of (\ref{vecmoddef}), one then takes $e^{S_0}$ times the Gaussian determinant from integrating the exponential of the quadratic terms in (\ref{phiTay}), with all the higher-order terms neglected.

All the indications are that the results of the above mean-field analysis are exact at $N\to\infty$, and they compare very well with the corresponding numerics \cite{PN}. And yet, why could the higher-order terms in (\ref{phiTay}) be neglected? Most naively, differentiating $S$ with respect to $\phi$ inserts powers of $\sqrt{\be}$, and since $\be=O(1/N)$, this amounts to manifest negative powers of $N$, which is reassuring. This is, however, not yet the whole story. The number of variables grows with $N$, and in the standard Feynman-like expansion while evaluating (\ref{vecmoddef}) with (\ref{phiTay}) substituted, one will have multiple sums over indices ranging from 1 to $N$. This will introduce positive powers of $N$ that will compete with the negative powers. (The ability of fluctuations of a large number of variables to compromise $1/N$ expansions is highlighted in a similar context in section 8.2 of \cite{Polyakov}.) More than that, if we repeatedly differentiate $S$ with respect to the same component of $\phi$, the resulting expression has $N$ terms, contributing further positive factors of $N$.

Due to all the above ingredients, the story with higher-order corrections to \cite{PN}, which we briefly sketch in Appendix~\ref{appPNexp}, appears somewhat bewildering. While the naive evaluation
within the Gaussian approximation produces the correct result at large $N$ that can be verified numerically, the higher-order corrections come with a swarm of positive and negative powers of $N$, and it is difficult to give a concise argument as to how exactly they are suppressed, and why they do not upset the Gaussian estimate (though it is known empirically that they do not). While one could attempt diagrammatic accounting of the powers of $N$ in these corrections, somewhat in the spirit of random tensor considerations \cite{randtens,randtensbook}, it is anything but easy.

Instead of attempting diagrammatic analysis of (\ref{vecmoddef}) with the expansion (\ref{phiTay}), we shall follow here a different strategy,
introducing a further set of auxiliary variables, somewhat akin to those recently used for analyzing graphs with prescribed degree sequences
\cite{Kawamoto,regcount,degseq}. In this way, conventional saddle points will emerge that control the large $N$ limit of (\ref{vecmoddef}),
where $N$ becomes a numerical parameter in the `action,' and the set of integration variables is $N$-independent. This will have other useful applications, besides elucidating the analytics behind the mean-field solution of \cite{PN}.

We first write (\ref{vecmoddef}) as
\beq
Z=\frac1{\pi^{N/2}}\int_{-\infty}^\infty \left(\prod_k 
\frac{d\phi_k\,e^{-\phi_k^2}}{\sqrt{1+e^{-2\al+4i\phi_k \sqrt{\be}}}}\right)
\prod_{k,l=1}^N\exp\left[\frac12\log\left(1+e^{-2\al+2i(\phi_k+\phi_l) \sqrt{\be}}\right)\right],
\eeq
and then expand the logarithm as $\log(1+x)=x-x^2/2+x^3/3+\cdots$ to obtain
\begin{align*}
Z&=\frac1{\pi^{N/2}}\int_{-\infty}^\infty \left(\prod_k 
\frac{d\phi_k\,e^{-\phi_k^2}}{\sqrt{1+e^{-2\al+4i\phi_k \sqrt{\be}}}}\right)
\prod_{k,l=1}^N\exp\left[-\sum_{J=1}^\infty\frac{(-1)^J}{2J} e^{J(-2\al+2i(\phi_k+\phi_l) \sqrt{\be})}\right]\nn\\
&=\frac1{\pi^{N/2}}\int_{-\infty}^\infty \left(\prod_k 
\frac{d\phi_k\,e^{-\phi_k^2}}{\sqrt{1+e^{-2\al+4i\phi_k \sqrt{\be}}}}\right)
\exp\left[-\sum_{J=1}^\infty\frac{(-1)^Je^{-2\al J}}{2J} \Big(\sum_k e^{2iJ\phi_k \sqrt{\be}}\Big)^{\!2\,}\right].
\end{align*}
An attractive feature of the last expression is that it can be completely factorized over $\phi_k$ at the cost of introducing one more Hubbard-Stratonovich transformation with respect to the variables $x_J$ that couple to $\sum_k e^{2iJ\phi_k \sqrt{\be}}$. In this way, one gets
\beq\label{xJHS}
Z=\frac1{\pi^{N/2}}\int_{-\infty}^\infty \left(\prod_{J=1}^\infty \frac{dx_J}{\sqrt{2\pi J}} \,e^{- x_J^2/2J}\right)\prod_k\int_{-\infty}^\infty \hspace{-3mm}d\phi_k
\frac{e^{-\phi_k^2}\exp\left[\sum_{J=1}^\infty i^{J+1}e^{-J\al}e^{2iJ \phi_k \sqrt{\be}}x_J/J\right]}{\sqrt{1+e^{-2\al+4i\phi_k \sqrt{\be}}}},
\eeq
or more suggestively, since all the $\phi_k$ integrals are identical to each other:
\begin{align}\label{Zx}
&Z=\int_{-\infty}^\infty \left(  \prod_{J=1}^\infty\frac{dx_J}{\sqrt{2\pi J}} \,e^{- x_J^2/2J}\right)\, e^{NS[x]},\\
&e^{S[x]}=\frac{1}{\sqrt{\pi}}\int_{-\infty}^\infty d\phi \frac{e^{-\phi^2}}{\sqrt{1+e^{-2\al+4 i\phi\sqrt{\be}}}}\exp\left[\sum_{J=1}^\infty\frac{i^{J+1}}{J} e^{-J\al}x_Je^{2iJ\phi\sqrt{\be}}\right].\label{Seff}
\end{align}
The integral in the last line gives an effective action $S[x]$, in terms of which one has to deal with a saddle point problem at large $N$, with $N$ playing no more of a role than providing a large saddle point parameter.


\section{The dense regime at leading order}\label{secPN}

We now turn to the dense regime of \cite{PN}, where $\al\sim 1$ and $\be\sim 1/N$, which we write as 
\beq
\be=\frac{B}{N}.
\eeq 
These scalings are necessary to ensure a nontrivial infinite $N$ limit with a finite mean connectivity (the fraction of all possible edges that are filled).

For the first-pass treatment in this section, as in \cite{PN}, we only keep the 
extensive part of the free energy, that is, contributions to $(\log Z)/N$ that survive at $N\to\infty$. For that, we have compute $e^S$ up to the order $O(1)$ at $N\to\infty$, so that $e^{NS}$ is computed correctly up to factors that stay finite at $N\to\infty$. In this way, we reproduce the solution of the model at the precision level of \cite{PN}, while the subleading corrections will be discussed in the next section.

For constructing a saddle point estimate of (\ref{Zx}-\ref{Seff}), it is important to identify the $N$-scaling of the variables responsible for the dominant contribution. In view of the mean-field picture of \cite{PN} that has been verified by comparisons with numerics, the field $\phi$ condenses to values of order $\sqrt{N}$. Since $x_J$ are Hubbard-Stratonovich conjugates of $\sum_k e^{iJ\phi_k \sqrt{\be}}$, they are expected to condense at values of order $N$, since the sum over $k$ consists of $N$ identical terms of order 1. We shall see that this regime $\phi\sim\sqrt{N}$, $x_J\sim N$ is indeed consistent and produces a saddle point estimate of $Z$ with controlled corrections.

Motivated by the above picture, we write $\phi=i\sqrt{BN}\phi_0+\ph$. We will choose $\phi_0$ to ensure that the integral over $\ph$ is dominated by $\ph$ of order 1. Then, neglecting all terms suppressed by $1/N$,
\begin{align}
e^{S[x]}=&\frac{e^{NB\phi_0^2}}{\sqrt{\pi\left(1+e^{-2\al- 4B\phi_0}\right)}}\int_{-\infty}^\infty d\varphi\,e^{-\ph^2-2i\sqrt{BN}\phi_0\ph}\nn\\
&\times\exp\left[\sum_{J=1}^\infty\frac{i^{J+1}}{J} e^{-J(\al+2B\phi_0)}x_J\left(1+2iJ\ph\sqrt{B/N}-2J^2\ph^2B/N\right)\right].
\label{action}
\end{align}
We then choose $\phi_0$ to ensure that the terms of order $O(\sqrt{N})$ that are linear in $\ph$ cancel out, which would guarantee that only values of $\ph$ of order 1 contribute to the integral:
\beq\label{phi0eq}
\phi_0=\sum_{J=1}^\infty {i^{J+1}} e^{-J(\al+2B\phi_0)}\frac{x_J}{N}.
\eeq
One is then left with an elementary Gaussian integral over $\ph$ that yields
\beq
e^{S[x]}=\frac{e^{NB\phi_0^2}}{\sqrt{1+e^{-2\al-4 B\phi_0}}}\exp\left[\sum_{J=1}^\infty\frac{i^{J+1}}{J} e^{-J(\al+2B\phi_0)}x_J\right]\left(1+{2B}\sum_{J=1}^\infty {i^{J+1}} Je^{-J(\al+2B\phi_0)}\frac{x_J}{N}\right)^{-1/2}.
\eeq
The corrections to this formula are of order $1/N$, which will give at most multiplicative contributions of order $O(1)$ in $e^{NS}$, and those cannot affect the extensive part of the free energy.

From the above, we write $S$ as
\begin{align}\label{Sdom}
&S=NS_{\mathrm{main}}+S_{\mathrm{corr}},\qquad S_{\mathrm{main}}\equiv B\phi_0^2+\sum_{J=1}^\infty\frac{i^{J+1}}{J} e^{-J(\al+2B\phi_0)}\frac{x_J}N, \\
&S_{\mathrm{corr}}\equiv -\frac12\log\left(1+{2B}\sum_{J=1}^\infty {i^{J+1}} Je^{-J(\al+2B\phi_0)}\frac{x_J}{N}\right)-\frac12\log\left(1+e^{-2\al-4 B\phi_0}\right).\label{Sextra}
\end{align}
where the leading terms of order $N$ have been separated out explicitly, and $\phi_0$ is a function of $x_J$ implicitly determined by (\ref{phi0eq}).
To construct a saddle point estimate of (\ref{Zx}), we look for stationary points of $NS[x]-\sum_Jx_J^2/2J$, determined at leading order in $N$ by
\beq
\label{sadd_est}
\frac{\del}{\del x_K}\left(N^2S_{\mathrm{main}}[x]-\sum_J \frac{x_J^2}{2J}\right)=0.
\eeq
With the dominant terms in $S$ collected in (\ref{Sdom}), this yields
$$
x_K/K=2{N^2B}\phi_0\frac{\del\phi_0}{\del x_K}+N\frac{i^{K+1}}{K}e^{-K(\al+2B\phi_0)}-2BN\frac{\del\phi_0}{\del x_K}\sum_{J=1}^\infty i^{J+1} e^{-J(\al+2B\phi_0)}x_J.
$$
The terms with $\del\phi_0/\del x_K$ cancel out in view of (\ref{phi0eq}), leaving the following saddle point configuration:
\beq\label{Xsddl}
x_K=X_K\equiv N i^{K+1} e^{-K(\al+2B\phi_0)}.
\eeq
Note that $X_K$ are consistently of order $N$. At this saddle point, (\ref{phi0eq}) becomes an explicit equation for $\phi_0$
\beq\label{phi0atsadd}
\phi_0=\sum_{J=1}^\infty {(-1)^{J+1}} e^{-J(2\al+4B\phi_0)}=\frac{1}{e^{2\al+4B\phi_0}+1},
\eeq
as in the mean-field solution of \cite{PN}. If we expand (\ref{Zx}) around the saddle point configuration $X_K$, the Gaussian integral over fluctuations is manifestly of order $O(1)$ at large $N$, as we shall see more explicitly in the next section. Then, nonvanishing contributions to the free energy per vertex
may only come from the saddle point exponential evaluated at $X_K$, that is
\beq
\frac{\log Z}{N}=-\sum_J \frac{X_J^2}{2JN} + NS_{\mathrm{main}}[X]+S_{\mathrm{corr}}[X].
\eeq
From (\ref{Sdom}), (\ref{Sextra}) and (\ref{Xsddl}), this is evaluated as
\beq\label{PNres}
\frac{\log Z}{N}={NB}\phi_0^2+\frac{N-1}2\log\left(1+e^{-2\al-4B\phi_0}\right)-\frac12\log\left[1+{2B}\phi_0(1-\phi_0)\right].
\eeq
where we have used the evident formulas
$$
\sum_{J=1}^\infty \frac{(-1)^{J+1}}{J}z^J=\log(1+z),\qquad \sum_{J=1}^\infty {(-1)^{J+1}}z^J=\frac{z}{1+z},\qquad  \sum_{J=1}^\infty J{(-1)^{J+1}}z^J=\frac{z}{(1+z)^2},
$$
and also equation (\ref{phi0eq}) to simplify the terms of order 1. The expression \label{PNres} is identical to the result of \cite{PN} up to a change of notation: what we call $\alpha$ here is $-B$ in \cite{PN}, and what we call $B$ is $-JN/(N-1)$ in \cite{PN}.


\section{The dense regime: subleading corrections}\label{sec1N}

To keep the story clean, we shall take the saddle point values deduced in the previous section as an input for identifying the relevant point of expansion, and restart the derivation
independently departing, once again, from the partition function (\ref{Zx}-\ref{Seff}). We will see that this results in an explicit controllable $1/N$ expansion that we will process in a matter-of-fact manner.

In (\ref{Zx}-\ref{Seff}) with $\beta=B/N$, we introduce the following variable redefinitions motivated by the previous section:
\begin{align}\label{phiexp}
&\phi \equiv i\sqrt{BN}\phi_0 + \varphi,\qquad \phi_0\left(e^{2\al+4B\phi_0}+1\right)\equiv 1,\\
&x_K \equiv X_K + \chi_K, \qquad X_K\equiv N i^{K+1} e^{-K(\al+2B\phi_0)}.
\label{Xexp}
\end{align}
To avoid clutter in the sums, we define the shorthand
\beq
 c_J\equiv\frac{X_J}{JN}=\frac{i^{J+1}}{J} e^{-J(\al+2B\phi_0)}.
\eeq

In the previous section, we computed $S$ up to the order $O(1)$, and then the free energy is computed up to the order $O(N)$, since $S$ enters the expression for $Z$ as $e^{NS}$ and one has to take a logarithm of $Z$. In order to compute the next correction, which is the nonextensive contribution to the free energy of order $O(1)$, one correspondingly has to compute $S$ up to the order $O(1/N)$. To do so, with the above notation, we rewrite
(\ref{Seff}) as
$$
e^{S[\chi]}=e^{NB\phi_0^2}\int_{-\infty}^\infty \hspace{-3mm}d\varphi\,\frac{e^{-\varphi^2-2i\sqrt{BN}\phi_0\varphi}}{\sqrt{\pi\left(1+e^{-2\al- 4B\phi_0 }e^{4i\ph\sqrt{B/N}}\right)}}\exp\left[\sum_{J=1}^\infty c_J(X_J + \chi_J)e^{2iJ\varphi\sqrt{B/N}}\right].
$$
We expand the denominator in the integrand as
$$
\left(1+ae^{b/\sqrt{N}}\right)^{-1/2} =  (1+a)^{-1/2}\left(1-\frac{a}{2(1+a)}\frac{b}{\sqrt{N}}+\frac{a^2-2a}{8(1+a)^2}\frac{b^2}{N}+\cdots\right),
$$
and, keeping in mind that $X_J\sim N$, expand the integrand up to order $1/N$. Importantly, terms involving $\varphi \sqrt{N}$ in the exponent cancel out because of our choice of $\phi_0$ and $X_J$, and one is left with the following Gaussian integral (where we omit odd powers of $\varphi$, which integrate to 0):
\beq
\begin{split}
e^{S[\chi]}=&\frac{e^{NB\phi_0^2+N\sum_J c_J^2J+\sum_J  c_J\chi_J}}{\sqrt{\pi\left(1+e^{-2\al- 4B\phi_0}\right)}}  \int_{-\infty}^\infty d\varphi \;e^{-\ph^2\left(1 + {2B}\sum_J  c_J^2J^3\right)}\Bigg\{ 1 + O(N^{-2})\\
&+\frac{1}{N}\Bigg[ \frac{2B\ph^2(2e^{2\al + 4B\phi_0}-1)}{(1+e^{2\al + 4B\phi_0})^2}- 2B\ph^2\left(\sumc J\chi_J\right)^2 - 2B\ph^2\sumc J^2 \chi_J \\
& +  \frac{4B\ph^2}{1+e^{2\al + 4B\phi_0}}\sumc  J\chi_J+ \frac{2B^2\ph^4}{3}\tsum c_J^2 J^5+\frac{8B^2\varphi^4}{3}\textstyle\sum_J c_J J \chi_J \textstyle\sum_Kc_K^2 K^4\\
&\hspace{2cm} -\frac{8B^2\ph^4}{3(1+e^{2\al + 4B\phi_0})}\,\tsum c_J^2 J^4   - \frac{8B^3\ph^6}{9}\left(\tsum c_J^2 J^4\right)^2\Bigg]\Bigg\}.
\end{split}
\label{phintexpand}
\eeq
We will often encounter the following sums, for which we introduce explicit notation:
\beq\label{sigdef}
\s_n \equiv\sum_{J=1}^{\infty} c_J^2  J^{n+1}= \sum \frac{(-1)^{J+1}}{J}\left(\frac{\phi_0}{1-\phi_0}\right)^{\!\! J}J^n  = \left(x\frac{d}{dx}\right)^{\!\! n}\log(1+x)\bigg|_{x=\frac{\phi_0}{1-\phi_0}}.
\eeq
Evaluating the $\ph$-integral in (\ref{phintexpand}) then yields
\beq\label{eS}
\begin{split}
e^{S[\chi]}=&\frac{e^{NB\phi_0^2+N\s_0+\sum_J  c_J\chi_J}}{\sqrt{(1+e^{-2\al- 4B\phi_0})(1+2B\s_2)}} \Bigg\{ 1+ O(N^{-2})\\
&+\frac{1}{N}\Bigg[ \frac{B(2e^{2\al + 4B\phi_0}-1)}{(1+e^{2\al + 4B\phi_0})^2(1+2B\s_2)}- \frac{B\left(\sumc J\chi_J\right)^2}{1+2B\s_2} - \frac{B\sum_J c_J  J^2 \chi_J}{1+2B\s_2} \\
&\hspace{2cm}+  \frac{2B\sum_J c_J  J\chi_J}{(1+e^{2\al + 4B\phi_0})(1+2B\s_2)}+ \frac{B^2\s_4}{2(1+2B\s_2)^2}+\frac{2B^2\sigma_3\sumc J\chi_J}{(1+2B\s_2)^2} \\
&\hspace{2cm} -\frac{2B^2\s_3}{(1+e^{2\al + 4B\phi_0})(1+2B\s_2)^2}   - \frac{5B^3\s_3^2}{3(1+2B\s_2)^3}\Bigg]\Bigg\}.
\end{split}
\eeq
Raising this expression to the power of $N$, so as to substitute it in (\ref{Zx}), and making use of $(1+x/N)^N \sim e^x$ at large $N$, we arrive at
\begin{align}
e^{NS[\chi]}=&\frac{e^{N^2B\phi_0^2+N^2\s_0+N\sum_J  c_J\chi_J}}{\left[(1+e^{-2\al- 4B\phi_0})(1+2B\s_2)\right]^{N/2}} \,\,\exp\Bigg[\frac{B(2e^{2\al + 4B\phi_0}-1)}{(1+e^{2\al + 4B\phi_0})^2(1+2B\s_2)}\\
&\hspace{1cm}+ \frac{B^2\s_4}{2(1+2B\s_2)^2} -\frac{2B^2\s_3}{(1+e^{2\al + 4B\phi_0})(1+2B\s_2)^2}   - \frac{5B^3\s_3^2}{3(1+2B\s_2)^3}\nn\\ 
&\hspace{-5mm}+\frac{2B^2\sigma_3\sumc J\chi_J}{(1+2B\s_2)^2}+  \frac{2B\sum_J c_J  J\chi_J}{(1+e^{2\al + 4B\phi_0})(1+2B\s_2)}- \frac{B\sum_J c_J J^2 \chi_J}{1+2B\s_2}+  - \frac{B\left(\sumc J\chi_J\right)^2}{1+2B\s_2}\Bigg],\nn
\end{align}
where all possible corrections to this expression are suppressed by powers of $1/N$.
Evaluating $\s_n$ in terms of $\phi_0$ from (\ref{sigdef}) -- the explicit formulas are tabulated in Appendix~\ref{appsigma} -- and also keeping in mind that $e^{2\al+4B\phi_0} = (1-\phi_0)/{\phi_0}$ and $1+e^{-2\al-4B\phi_0} = {1}/(1-\phi_0)$, we obtain
\begin{align}\label{eNS}
e^{NS[\chi]}=&\frac{e^{N^2B\phi_0^2}(1-\phi_0)^{-N^2+N/2}e^{N\sum_J  c_J\chi_J}}{[1+2B\phi_0(1-\phi_0)]^{N/2}} \,\exp\Bigg[\frac{B\phi_0(2-3\phi_0)}{1 + {2B}\phi_0(1-\phi_0)}\\
& \hspace{-1.2cm} - \frac{2B^2\phi_0^2(1-\phi_0)(1-2\phi_0)}{[1 + {2B}\phi_0(1-\phi_0)]^2}+ \frac{B^2 \phi_0(1-\phi_0)(1-6\phi_0+6\phi_0^2)}{2[1 + {2B}\phi_0(1-\phi_0)]^2}  - \frac{5B^3\phi_0^2(1-\phi_0)^2(1-2\phi_0)^2}{3[1 + {2B}\phi_0(1-\phi_0)]^3}\nn \\
&\hspace{-5mm}+\frac{2B\phi_0[1+B(1-\phi_0)]\sumc J \chi_J}{[1 + {2B}\phi_0(1-\phi_0)]^2} - \frac{B\left(\sum_J c_J J\chi_J\right)^2}{1 + {2B}\phi_0(1-\phi_0)} - \frac{B\,\sum_J c_J J^2 \chi_J}{1 + {2B}\phi_0(1-\phi_0)}\Bigg].\nn
\end{align}
From this,
\begin{align}\label{Zintexponent}
&-\sum_{J}\frac{x_J^2}{2J} + NS[x] = N^2B\phi_0^2 - \frac{N(N-1)}{2}\log(1-\phi_0) -\frac{N}{2}\log[1+2B \phi_0(1-\phi_0)]\\
&\hspace{5mm}+\frac{B\phi_0(2-3\phi_0)}{1 + {2B}\phi_0(1-\phi_0)} +\frac{B^2\phi_0(1-\phi_0)(1-10\phi_0+14\phi_0^2) }{2[1+2B \phi_0(1-\phi_0)]^{2}}- \frac{5B^3\phi_0^2(1-\phi_0)^2(1-2\phi_0)^2}{3[1 + {2B}\phi_0(1-\phi_0)]^3}\nn\\
&\hspace{1cm} +\frac{2B\phi_0[1+B(1-\phi_0)]\sumc J \chi_J}{[1 + {2B}\phi_0(1-\phi_0)]^2} - \frac{B\,\sum_J c_J J^2 \chi_J}{1 + {2B}\phi_0(1-\phi_0)}- \frac{B\left(\sumc J\chi_J\right)^2}{1 + {2B}\phi_0(1-\phi_0)}.\nn
\end{align}
Importantly, the term involving $N\sum_J c_J\chi_J$ in the first line of (\ref{eNS}) cancels out in (\ref{Zintexponent}). This leaves in (\ref{Zx}) a Gaussian integral over $\chi_J$ where all integration variables are only allowed to take values of order 1, leading to an explicit finite result. This reflects the suitability of our parametrization
in (\ref{phiexp}-\ref{Xexp}). Had we chosen a different expansion point, large terms of the form $N\chi_J$ would have compromised the usefulness of evaluating the Gaussian integrals at this stage, as they would induce rearrangements in the $1/N$ corrections.

To evaluate (\ref{Zx}), it remains to integrate $e^{-\sum_{J}{x_J^2}/{2J} + NS[x]}$ over $\chi_J$. To do so, we employ the following Hubbard-Stratonovich transformation for the exponential of the last term in (\ref{Zintexponent}):
\beq\label{HSy}
\exp\left[-\frac{B\left(\sumc J\chi_J\right)^2}{1+2B \phi_0(1-\phi_0)}\right] = \int_{-\infty}^\infty dy \frac{e^{-y^2}}{\sqrt\pi}\exp\left[2yi\sqrt{\frac{B}{1+2B \phi_0(1-\phi_0)}}\,\,\sumc J\chi_J\right].
\eeq
It is now straightforward, even if somewhat laborious, to evaluate the remaining Gaussian integral over $\chi_J$ where the quantities $\s_n$ defined in (\ref{sigdef}) appear once again, and can thereafter be explicitly expressed through $\phi_0$. The result is
\begin{align}
&\prod_{J=1}^\infty\int_{-\infty}^\infty  \frac{d\chi_J\,e^{-{\chi_J^2}/{2J}}}{\sqrt{2\pi J}}\exp{\textstyle\left[c_J J\left( \frac{2B\phi_0[1+B(1-\phi_0)]}{[1 + {2B}\phi_0(1-\phi_0)]^2}   - \frac{JB}{1 + {2B}\phi_0(1-\phi_0)}+ 2yi \sqrt{\frac{B}{1+2B \phi_0(1-\phi_0)}}\right)\chi_J \right]}\nn\\
&\hspace{1cm}=\exp\Big[{\textstyle\frac{B^2\phi_0(1-\phi_0)(1-10\phi_0+18\phi_0^2)+8B^3\phi_0^3(1-\phi_0)^2(5\phi_0-2)+8B^4\phi_0^4(1-\phi_0)^3(3\phi_0-1)}{2[1 + {2B}\phi_0(1-\phi_0)]^4}}\label{chiint}\\
&\hspace{2.5cm}{\textstyle +2yi \left(\frac{B}{1+2B \phi_0(1-\phi_0)}\right)^{3/2}\frac{\phi_0(1-\phi_0)(4\phi_0-1)+4B\phi_0^3(1-\phi_0)^2}{1 + {2B}\phi_0(1-\phi_0)} -\frac{2y^2B\phi_0(1-\phi_0)}{1+2B \phi_0(1-\phi_0)}}\Big].\nn
\end{align}
We have once again used the expressions for sums of the form $\sum_J c_J^2 J^{n+1}$ given in Appendix~\ref{appsigma}.
To undo the Hubbard-Stratonovich transformation in (\ref{HSy}), we need to multiply the last line with $e^{-y^2}/\sqrt{\pi}$ and integrate over $y$.
This yields:
\begin{align}
&\int_{-\infty}^\infty \frac{dy}{\sqrt\pi}\exp\Bigg[{\textstyle -y^2\Bigg(1+\frac{2B\s_2}{1+2B \phi_0(1-\phi_0)}\Bigg)+2yi \left(\frac{B}{1+2B \phi_0(1-\phi_0)}\right)^{3/2}\frac{\phi_0(1-\phi_0)(4\phi_0-1)+4B^2\phi_0^3(1-\phi_0)^2}{1 + {2B}\phi_0(1-\phi_0)}}\Big]\nn\\
&\hspace{2mm}=\exp\left\{-\frac{B^3\phi_0^2(1-\phi_0)^2[4\phi_0-1+4B\phi_0^2(1-\phi_0)]^2}{[1+2B \phi_0(1-\phi_0)]^4[1+4B \phi_0(1-\phi_0)]}\right\}\left(\frac{1+4B\phi_0(1-\phi_0)}{1+2B \phi_0(1-\phi_0)}\right)^{-1/2}.
\label{yintres}
\end{align}
Finally, we need to gather all the contributions to $Z$ of (\ref{Zx}), consisting of the $\chi$-independent part of (\ref{Zintexponent}), the second $y$-independent line of (\ref{chiint}), and (\ref{yintres}). This leaves for the free energy
\begin{align}\label{Forder1}
&-F \equiv \log Z = N^2B\phi_0^2 - \frac{N(N-1)}{2} \log(1-\phi_0)-\frac{N}{2}\log\left[1+2\Delta\right]\\\
&\hspace{1cm}+\frac{2\Delta-B\phi_0^2}{1 + 2\Delta}  -\frac{5B\Delta^2(1-2\phi_0)^2}{3(1 + 2\Delta)^3}-\frac{B\Delta^2[4\phi_0(1+\Delta)-1]^2}{(1+2\Delta)^4(1+4\Delta)} - \frac12\log\frac{1+4\Delta}{1+2\Delta}\nn\\
&\hspace{1cm}+\frac{B\Delta\big[1-10\phi_0+16\phi_0^2+2\Delta(1-14\phi_0+24\phi_0^2)+2\Delta^2(1-12\phi_0+20\phi_0^2)\big]}{(1 + 2\Delta)^4}.\nn
\end{align}
with $\Delta\equiv B \phi_0(1-\phi_0)$ introduced for compactness.
The first line reflects the mean-field result of \cite{PN}, while the subsequent two lines give the first nontrivial correction that the novel representations developed here allowed us to derive.

As the computations leading to (\ref{Forder1}) are rather convoluted, it is important to implement some independent checks. One such check is provided by the particle-hole duality respected by the two-star model: in the original partition function (\ref{2stardef}), one can change the summation variable from the adjacency matrix $A$ to the `inverted' adjacency matrix, where all filled edges are replaced by empty ones and vice versa. This manifestly relates partition functions, and hence free energies, at two different values of $\alpha$ at any given $N$. As (\ref{Forder1}) is expressed through $\phi_0(\al,B)$ rather than $\al$, one has to deal with the corresponding transformation for $\phi_0$ that maps it to $1-\phi_0+O(1/N)$. The subleading $1/N$ corrections in this transformation mix the different orders of $N$ in (\ref{Forder1}) and since the different contributions must cancel out in the end, this provides a nontrivial check of the subleading corrections in the last two lines of (\ref{Forder1}).
Our formula passes this test. We provide an explicit implementation in Appendix~\ref{appdual}.

Further validation of (\ref{Forder1}) is provided by comparisons with numerical Monte Carlo sampling of the two-star model. We follow a strategy similar to what
we have previously employed in \cite{birth} for more complicated related models, reviewed in Appendix~\ref{appMC}. In our comparisons between analytics and numerics, we focus on the degree variance. In terms of the free energy $F\equiv -\log Z$, the averages of the mean degree $\langle k\rangle \equiv \sum_j \langle k_j\rangle/N=\langle k_1\rangle$ and mean degree squared
$\langle k^2\rangle \equiv \sum_j \langle k_j^2\rangle/N=\langle k_1^2\rangle$ are expressed as
\beq
\left<k\right> = \frac{1}{N}\frac{\del F}{\del \al},\qquad
\left<k^2\right> = \frac{\del F}{\del B}.
\eeq
To compute these derivatives, one also needs the derivatives of $\phi_0$ with respect to $\al$ and $B$:
\beq
\frac{\del\phi_0}{\del\al} = -\frac{2\phi_0(1-\phi_0)}{1 + 4B\phi_0(1-\phi_0)},\qquad
\frac{\del\phi_0}{\del B} =  -\frac{4\phi_0^2(1-\phi_0)}{1 + 4B\phi_0(1-\phi_0)}.
\eeq
Then, the variance is
\beq
V\equiv\left<k^2\right> -\left<k\right>=\frac{\del F}{\del B} - \left(\frac{1}{N}\frac{\del F}{\del \al}\right)^2.
\eeq
We separate out the contribution due to the extensive part of free energy, given by the first line of (\ref{Forder1}):
\begin{equation}
-F_0 \equiv N^2B\phi_0^2 - \frac{N(N-1)}{2} \log(1-\phi_0)-\frac{N}{2}\log\left[1+2B \phi_0(1-\phi_0)\right].
\end{equation}
The contribution to the degree variance $V$ coming from $F_0$ is
\begin{equation}\label{F0var}
\frac{\del F_0}{\del B} - \left(\frac{1}{N}\frac{\del F_0}{\del \al}\right)^2 = N\frac{\phi_0(1-\phi_0)}{1+2\Delta} - \frac{\phi_0^2[2B\phi_0(\phi_0-2)+2B+1]^2}{(1+2\Delta)^2(1+4\Delta)^2},
\end{equation}
with $\Delta=B\phi_0(1-\phi_0)$.
The leading order variance 
\beq\label{defV0}
V_0\equiv N\frac{\phi_0(1-\phi_0)}{1+2\Delta} 
\eeq
matches the one derived in \cite{PN}. We are keeping the subleading order in (\ref{F0var}) as it will combine with the higher-order corrections. The remaining nonextensive piece of the free energy (\ref{Forder1}) is
\begin{align}
-F_{corr} \equiv&\frac{2\Delta-B\phi_0^2}{1 + 2\Delta}  -\frac{5B\Delta^2(1-2\phi_0)^2}{3(1 + 2\Delta)^3}-\frac{B\Delta^2[4\phi_0(1+\Delta)-1]^2}{(1+2\Delta)^4(1+4\Delta)} - \frac12\log\frac{1+4\Delta}{1+2\Delta}\nn\\
&+\frac{B\Delta\big[1-10\phi_0+16\phi_0^2+2\Delta(1-14\phi_0+24\phi_0^2)+2\Delta^2(1-12\phi_0+20\phi_0^2)\big]}{(1 + 2\Delta)^4}.
\label{eq:F_corr}
\end{align}
Since this expression is bulky, it is more practical to use symbolic computation software for handling its derivatives, and we rely on  SymPy \cite{sympy} for this purpose,
followed by evaluating the resulting expressions for each set of parameters. To verify (\ref{Forder1}) numerically, we then extract point-by-point the degree variance from Monte Carlo simulations, with $V_0$ subtracted, and compare it with
\begin{equation}\label{defVcorr}
V_{corr}\equiv  \frac{\del F_{corr}}{\del B} - \left(\frac{1}{N}\frac{\del F_{corr}}{\del \al}\right)^2- \frac{\phi_0^2[2B\phi_0(\phi_0-2)+2B+1]^2}{(1+2\Delta)^2(1+4\Delta)^2}.
\end{equation}

\begin{figure}[t]
\centering
\includegraphics[width = 0.49\linewidth]{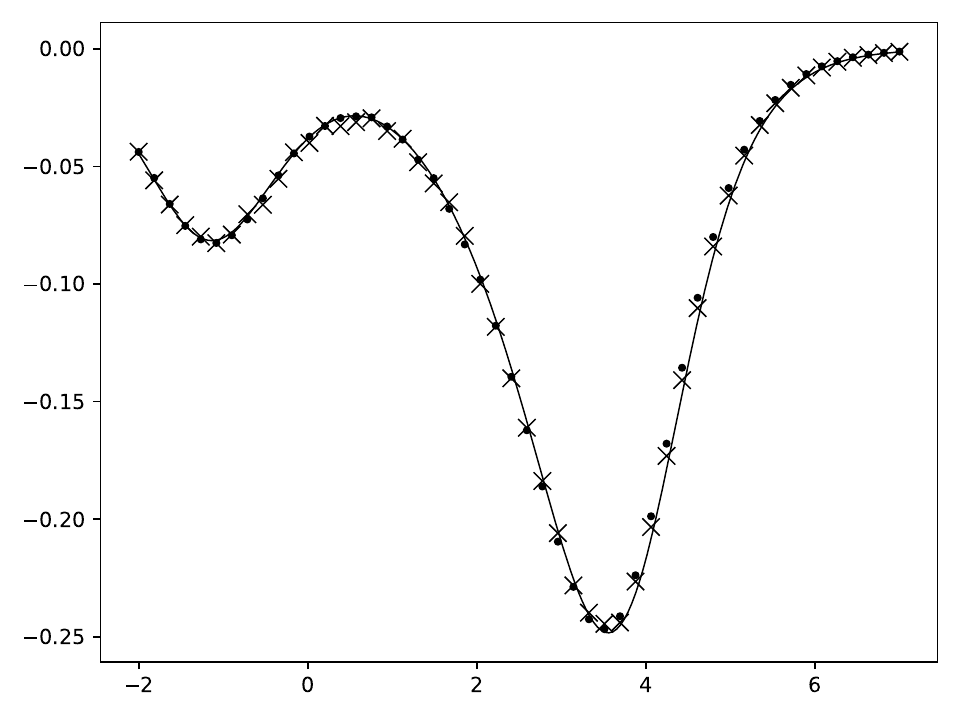}\hspace{2mm}\includegraphics[width = 0.49\linewidth]{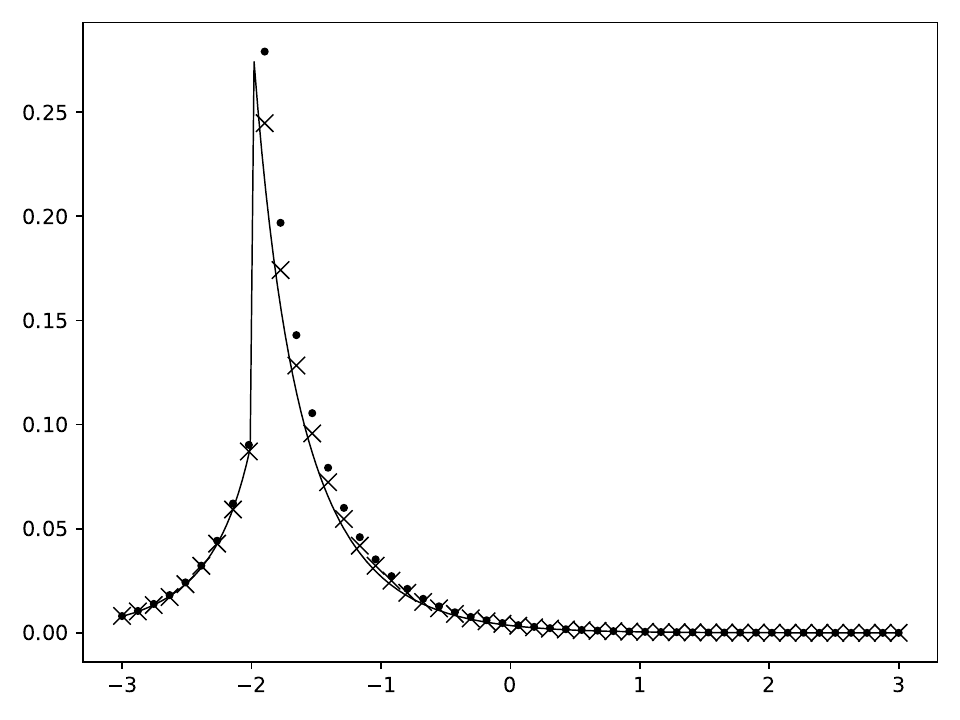}
\begin{picture}(0,0)
\put(-438,152){$\delta V$}
\put(-338,0){$\al$}
\put(-345,152){$B=1.3$}
\put(-205,152){$\delta V$}
\put(-100,0){$\al$}
\put(-112,152){$B=-2$}
\end{picture}\vspace{2mm}
\caption{The difference $\delta V$ between degree variance and the leading order prediction $V_0$ given by (\ref{defV0}) as a function of $\al$. The crosses represent numerical measurements from Monte Carlo sampling for graphs with $N=200$ vertices, the dots are the same for $N=50$ vertices, deviating from the analytic curve slightly stronger because of higher $1/N$ corrections, while the solid lines are the analytic prediction (\ref{defVcorr}). Two different values are presented: $B=1.3$ on the left and $B=-2$ on the right.}
\label{figvar}
\end{figure}
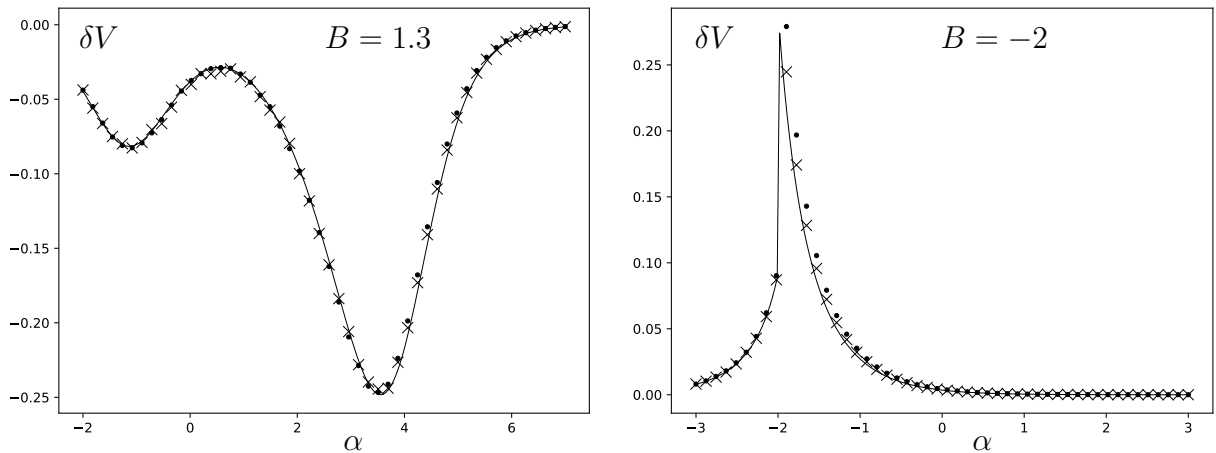

In Fig.~\ref{figvar}, we present the results of this comparison, showing excellent agreement. We opt for the moderate number of vertices $N=200$ as it allows us to keep the higher order $1/N$ corrections to the analytics sufficiently small, but at the same time the Monte Carlo equilibration for such moderately sized graphs happens sufficiently fast to attain a very high precision. This high precision is necessary since we are measuring a small subleading contribution to the variance. (We additionally display the results of numerical simulations for $N=50$.)

We have been dealing above with the dominant $1/N$ correction to the mean-field result of \cite{PN}.
If one is to compute further $1/N$ corrections to the free energy, the process is completely algorithmic, even if it will become more and more laborious with each next order. Indeed, higher-order corrections will manifest themselves as contributions of order $1/N^2$ and smaller in the formula for $e^S$ on top of what is written explicitly in
(\ref{eS}). All of these contributions are furthermore polynomial in $\chi_J$. Then, when writing  $e^{-\sum_{J}{x_J^2}/{2J} + NS[x]}$ in (\ref{Zx}), these corrections
will give contributions suppressed by $1/N$ or more in the exponent, all of which can be re-expanded as additive contributions polynomial in $\chi_J$. Thus, one will end up with the same Gaussian integrals as in our treatment above, except that they will have extra polynomial insertions in terms of $\chi_J$ suppressed by powers of $N$.
All such Gaussian integrals with polynomial insertions can be evaluated using the standard formulas, leaving behind an explicit series in powers of $1/N$ that will correct
our result in (\ref{Forder1}).


\section{The sparse regime}\label{secsparse}

We now turn to the sparse regime, 
\beq\label{scalesparse}
\al=\frac12\log\frac{N}c,\qquad \be=O(1).
\eeq
In this case, the $J$-sum in (\ref{Seff}) works very differently from the dense case considered above, since $e^{-J\al}$ turns into powers of $c/N$, introducing stronger and stronger $1/N$ suppression in contributions from higher $J$. This leads to considerable simplifications.

To identify the relevant scalings of $x_J$ in terms of powers of $N$ explicitly, consider the $1/N$ expansion of $e^{S}$ given by (\ref{Seff}), similar to what has been done in the previous sections, but with the new large-$N$ scalings (\ref{scalesparse}) of $\alpha$ and $\be$:
\begin{align*}
e^{S[x]}=&\frac{1}{\sqrt{\pi}}\int_{-\infty}^\infty d\phi \frac{e^{-\phi^2}}{\sqrt{1+ce^{4 i\phi\sqrt{\be}}/N}}\exp\Bigg[-\sqrt{\frac{c}N}x_1e^{2i\phi\sqrt{\be}}-\frac{ic}{2N}x_2e^{4i\phi\sqrt{\be}}\\
&\hspace{8cm}+\frac13\left(\frac{c}N\right)^{3/2}x_3\,e^{6i\phi\sqrt{\be}}+\cdots\Bigg].
\end{align*}
From this, we see that it is self-consistent to assume that $x_{J\ge 2}=O(1)$. Indeed, if that is so, the leading contribution to $e^S$ from $x_{J\ge 2}$ scales as $N^{-J/2}$. When raised to the power $N$ to obtain $e^{NS}$ in (\ref{Zx}), this will yield $\exp[O(N^{1-J/2})]$, since $[1+O(N^{-\gamma})]^N\sim e^{O(N^{1-\gamma})}$. Thus, for $J\ge 3$, the contribution of $e^{NS}$ in (\ref{Zx}) is completely negligible at large $N$, and one is left with an empty Gaussian integral of $e^{-x_J^2/2J}$, so that the fluctuations of $x_{J\ge 3}$ are of order 1 as assumed. For $x_2$, $e^{NS}$ gets a contribution of order 1 that should be integrated together with the Gaussian factor $e^{-x_2^2/4}$ in (\ref{Zx}). This integral will just produce a factor of order 1 in the partition function, and hence a nonextensive contribution to the free energy that does not alter the expectation values of the vertex degrees and their squares at large $N$, so we can ignore it. The fluctuations of $x_2$ are again of order 1 as assumed. The only contribution where the structure is different, and it defines the end result, is that of $x_1$. Indeed, assuming that $x_1$ is of order 1 is inconsistent, since that would have yielded $e^{NS}$ of the form $e^{O(\sqrt{N})}$, and the Gaussian factor $e^{-x_1^2/2}$ would not have been able to keep the fluctuations at order 1 as assumed. The consistent assumption is that $x_1$ is of order $\sqrt{N}$, since in that case, $e^S$ is of order 1, and hence $e^{NS}$ is of the form $e^{O(N)}$, while the Gaussian factor $e^{-x_1^2/2}$ is of the same order, so that the two factors balance each other self-consistently producing an explicit saddle point, as we shall see immediately.

From the above estimates, for the leading order analysis, $S$ will only depend on $x_1$ at large $N$, and to incorporate the relevant scaling of $x_1$, we write $x_1\equiv-x\,\sqrt{N/c\,}$. Then,
\beq\label{sparseZS}
Z= \sqrt{ \frac{N}{2\pi c}}\int_{-\infty}^\infty dx \,e^{N[S(x,\be)-x^2/2c]},\quad
S(x,\be)\equiv\log \frac{1}{\sqrt{\pi}}\int_{-\infty}^\infty d\phi\, \exp\left[-\phi^2+x\, e^{2i\phi\sqrt{\be}}\right].
\eeq
The first integral has an explicit saddle point structure at large $N$, and the saddle point is at $x=X$ satisfying
\beq\label{sddlsparse}
X=c\,\frac{\del S}{\del x}\Bigg|_{x=X}.
\eeq
The free energy per vertex $f$ satisfies
\beq
-f\equiv \frac{\log Z}{N}=S(X,\be)-X^2/2c,
\eeq
where $X$ is the solution of (\ref{sddlsparse}), and we have only kept the contributions nonvanishing at large $N$. It is convenient to represent $e^{S(x)}$ as a series
\beq
e^{S(x,\be)}=\frac{1}{\sqrt{\pi}}\int_{-\infty}^\infty d\phi\, e^{-\phi^2}\sum_{d=0}^\infty\frac{x^d}{d!} e^{2id\phi\sqrt{\be}}=\sum_{d=0}^\infty\frac{x^d}{d!} e^{-\beta d^2}.
\eeq
From (\ref{sddlsparse}),
\beq\label{Xser}
X=c\,\frac{\sum_{d=1}^\infty X^{d-1}e^{-\beta d^2}/(d-1)!}{\sum_{d=0}^\infty X^d e^{-\beta d^2}/d!}.
\eeq
The mean degree is
\beq
\langle k \rangle\equiv \frac{\del f}{\del\al}=\frac{dc}{d\al}\frac{\del f}{\del c}=-2c\frac{\del f}{\del c}=\frac{X^2}{c}+2c\frac{\del X}{\del c}\frac{\del}{\del X}\left[S(X,\be)-X^2/2c\right].
\eeq
The last term vanishes by the saddle point equation, giving
\beq\label{kXc}
\langle k \rangle=\frac{X^2}{c}.
\eeq
At the same time,
\beq
\langle k^2 \rangle\equiv \frac{\del f}{\del\be}=-\frac{\del S(X,\be)}{\del\be}-\frac{\del X}{\del \be}\frac{\del}{\del X}\left[S(X,\be)-X^2/2c\right].
\eeq
Again, due to the saddle point equation,
\beq\label{k2eq}
\langle k^2 \rangle=-\frac{\del S(X,\be)}{\del\be}=\frac{\sum_{d=1}^\infty d^2 X^d e^{-\beta d^2}/d!}{\sum_{d=0}^\infty X^d e^{-\beta d^2}/d!}.
\eeq
Expressing $X$ from (\ref{kXc}) and substituting it into (\ref{Xser}), multiplied by $X$, and into (\ref{k2eq}), we obtain:
\begin{align}\label{avk}
\langle k \rangle&=\frac{\sum_{d=1}^\infty d\,c^{d/2}\,\langle k \rangle^{d/2}e^{-\beta d^2}/d!}{\sum_{d=0}^\infty c^{d/2}\,\langle k \rangle^{d/2} e^{-\beta d^2}/d!},\\
\langle k^2 \rangle&=\frac{\sum_{d=1}^\infty d^2 c^{d/2}\,\langle k \rangle^{d/2}e^{-\beta d^2}/d!}{\sum_{d=0}^\infty c^{d/2}\,\langle k \rangle^{d/2} e^{-\beta d^2}/d!}.\label{avk2}
\end{align}
This is identical to the Annibale-Courtney solution of \cite{AC}, but now derived using only elementary integrals, and in a few lines rather than a few pages.
At any given $c$ and $\be$, (\ref{avk}) has to be solved as a nonlinear equation for $\langle k\rangle$, whereupon $\langle k^2\rangle$ is computed directly from (\ref{avk2}).


\section{The sparse regime: condensation of the master variable}\label{seccond}

As we have seen in the previous section, the solution of the two-star model in the sparse regime is determined by a saddle point structure in terms of $x_1$. The Park-Newman variables $\phi_k$, on the other hand, do not condense to definite values, since there is a saddle point structure in the integral defining $Z$ in (\ref{sparseZS}), but not in the integral defining $S$. Likewise, the vertex degrees do not condense, since both mean degree and degree fluctuations are of order 1.

To develop a physical intuition for the condensation of $x_1$, it is natural to turn to its Hubbard-Stratonovich conjugate $\sum_j e^{2i\phi_j\sqrt{\beta}}/N$.
We can compute the expectation value of this observable by inserting it into the partition function (\ref{Zphi}):
\beq\label{expexpect}
\frac1N\left<{\textstyle \sum_j\,} e^{2i\phi_j\sqrt{\beta}}\right>=\frac1{\pi^{N/2}NZ}\int_{-\infty}^\infty [d\phi_n]\,\sum_j e^{2i\phi_j\sqrt{\beta}}\,e^{-\sum_k\phi_k^2}\prod_{k<l}\left(1+e^{-2\al+2i(\phi_k+\phi_l) \sqrt{\be}}\right).
\eeq
This expression can be processed in two different ways. First, we can introduce the $x_J$ variables, pushing the structure toward (\ref{Zx}-\ref{Seff}), and then apply the saddle point analysis of section~\ref{secsparse}. The only difference due to the new factor of $\sum_j e^{2i\phi_j\sqrt{\beta}}$ is that we should multiply the integrand on the right-hand side of (\ref{xJHS}) with $-e^\alpha x_1$, since with this extra factor, integrating out $x_1$ precisely reproduces $\sum_j e^{2i\phi_j\sqrt{\beta}}$. In this way,
\beq\label{expexpx}
\frac1N\left<{\textstyle \sum_j\,} e^{2i\phi_j\sqrt{\beta}}\right>=-\frac{e^\al}{NZ}\int_{-\infty}^\infty \left(  \prod_{J=1}^\infty\frac{dx_J}{\sqrt{2\pi J}} \,e^{- x_J^2/2J}\right)\, x_1\,e^{NS[x]}=\frac{X}c,
\eeq
where we have used $x_1\equiv-x\,\sqrt{N/c\,}$ and the fact that $x$ condenses to the definite value $X$ determined by (\ref{sddlsparse}), with fluctuations suppressed by $1/\sqrt{N}$ in view of the saddle point considerations of the previous section.
Alternatively, we can take (\ref{expexpect}) and return to the original adjacency matrix variables defining the two-star model.
We thus undo the step leading from (\ref{ZAphi}) to (\ref{Zphi}) to obtain
\begin{align}\label{expexpA}
\frac1N\left<{\textstyle \sum_j\,} e^{2i\phi_j\sqrt{\beta}}\right>&= \frac1{\pi^{N/2}NZ}\sum_{A}\sum_j \int_{-\infty}^\infty [d\phi_n]\,e^{2i\phi_j\sqrt{\beta}}\,e^{-\sum_l\phi_l^2}e^{\sum_{kl} A_{kl}(-\al+2i\phi_l \sqrt{\be})},\nn\\
&=\frac1{NZ}\sum_{A}\left( \sum_je^{-\beta(2k_j+1)}\right) e^{-H[A]}=\frac{e^{-\beta}}N\left<{\textstyle \sum_j\,} e^{-2\beta k_j}\right>,
\end{align}
where $H[A]$ is the original two-star Hamiltonian (\ref{2stardef}) and $k_j\equiv\sum_l A_{jl}$ are the  vertex degrees. 
Note that all vertices are equivalent so we can write $\left<{\textstyle \sum_j\,} e^{-2\beta k_j}\right>/N$ as simply $\langle e^{-2\beta k}\rangle$. From (\ref{expexpx}) and (\ref{expexpA}), we then obtain
\beq\label{Xcond}
X=ce^{-\be}\langle e^{-2\beta k}\rangle.
\eeq
In other words, the condensation value of the master variable $x_1$ that controls the sparse regime of the two-star model is proportional to the expectation value of a specific exponential function of the vertex degree in terms of the original random graph variables.

As a byproduct of these considerations, we arrive at the following curious thermodynamic relation: from (\ref{Xcond}) and (\ref{kXc}), 
\beq
\langle e^{-2\beta k}\rangle=\,e^{\beta}\,\sqrt{\frac{\langle k\rangle}{c}}.
\eeq
There is thus a relation, at large $N$ and in thermodynamic equilibrium, between the expectation values of degrees and their exponentials.


\section{Conclusions}

We have revisited solutions of the two-star random graph model in the thermodynamic limit, in both dense and sparse regime. Previous approaches relied
on the variables $\phi_k$ dating back to \cite{PN}, which are Hubbard-Stratonovich conjugates of the vertex degrees. We pointed out that considerable
empowerment of the formalism results from introducing further variables $x_J$ as in (\ref{xJHS}). These variables can be thought of as Hubbard-Stratonovich conjugates
of $\sum_k e^{2iJ\phi_k\sqrt{\be}}/N$. The usage of such exponentials of the fields is ubiquitous in conformal field theories \cite{yellow} -- see \cite{199} for a recent discussion -- but to the best of our knowledge they have not appeared previously in studies of random graphs.

In the dense regime, it is known that the variables $\phi_k$ condense themselves to definite values at large $N$. It is predictable that all variables $x_J$ condense to definite values as well, with small fluctuations. Our representation provides for effective control over these fluctuations, with a straightforward bookkeeping of the $1/N$ factors, which is challenging in the language of $\phi_k$. We thus manage to compute the nonextensive part of the free energy in the dense regime. In the sparse regime,
the situation is more peculiar: the variables $\phi_k$ do not condense -- the representation for $S$ in (\ref{sparseZS}) does not have a large $N$ saddle point structure -- but the variables $x_J$ do, leading to a simple saddle point calculation of the free energy. This calculation is phrased entirely in terms of elementary one-dimensional integrals.
The condensation value of the $x$-variable determining the large $N$ behavior in the sparse regime can be expressed in terms of the original graph geometry as the expectation value of a certain exponential of the vertex degree given by (\ref{Xcond}).

We schematically summarize the interplay between our work and the previous treatments in \cite{PN} and \cite{AC} in the following table:
\begin{center}
\begin{tabular}{|m{2.8cm}|c|c|c|}
\hline
&\small Park-Newman \cite{PN}&\small Annibale-Courtney \cite{AC}&\small current treatment\\\hline
\small Mean-field \newline dense regime&\checkmark&&\checkmark\\\hline
\small Systematic\newline \small $1/N$ corrections&&&\checkmark\\\hline
\small Sparse regime\rule{0mm}{6mm}\vspace{2mm}&&\checkmark&\checkmark\\\hline
\small Functional\newline integrals avoided&\checkmark&&\checkmark\\\hline
\end{tabular}
\end{center}
(We stress that this table only shows the aspects of \cite{PN,AC} relevant for the considerations here, while those papers additionally consider many other important questions, in particular, in relation to the phenomenology of phase transitions in two-star graphs.)

Besides elucidating the analytic structure of the two-star model, we hope that the techniques presented here will usefully transfer to further related settings.
A number of interesting extensions of the two-star model can be seen in the literature \cite{altstar,corr,schelling,pairwise,anotherdiego}, and they provide an attractive
avenue for exploration in this regard.

\section*{Acknowledgments}

We thank Sergei Nechaev for valuable discussions on closely related topics. OE is supported by the  C2F program at Chulalongkorn University and by NSRF via grant number B41G680029.

\appendix

\section{Naive expansion around the Park-Newman solution}\label{appPNexp}

To illustrate the power of our formalism, we would like to briefly contrast it with naive expansion around the Park-Newman solution \cite{PN}. In our notation, the representation for the partition function used for obtaining that solution is given by (\ref{vecmoddef}), which we reproduce here for convenience
\beq\label{vecmoddefapp}
Z=\frac1{\pi^{N/2}}\int_{-\infty}^\infty d\phi_1\ldots d\phi_N\,e^{-\sum_k\phi_k^2\,+\,S[\phi]},\quad
S[\phi]\equiv\sum_{k<l}\log\left(1+e^{-2\al+2i(\phi_k+\phi_l) \sqrt{B/N}}\right).
\eeq

To derive the mean-field solution, one assumes that the dominant configuration is site-independent, $\phi_k=i\sqrt{BN}\phi_0$, where $\phi_0$ satisfies the same equation we derived in section~\ref{secPN}:
\beq\label{phi0app}
\phi_0=\frac{1}{e^{2\al+4B\phi_0}+1}.
\eeq
One then expands as $\phi_k=i\sqrt{BN}\phi_0+\ph_k$ and writes
$$
-\sum_k\phi_k^2+S[\phi]=S_0+\sum_k \ph_k^2 +\frac12\sum_{kl}\frac{\del^2 S}{\del\phi_k\del\phi_l}\ph_k\ph_l+\frac16\sum_{klm}\frac{\del^3 S}{\del\phi_k\del\phi_l\del\phi_m}\ph_l\ph_l\ph_m+\cdots.
$$
Equation (\ref{phi0app}) ensures that there are no terms linear in $\ph_k$ in this expansion.
Then, the exponential of all of these terms starting with the cubic one should be expanded in a Taylor series, so that a Gaussian integral over $\ph_k$ emerges from (\ref{vecmoddefapp}) with an infinite series of polynomial insertions.

To get a sense of how this expansion works, we write out the general formulas for the derivative tensors 
\begin{equation}
\frac{\del^nS}{\del\phi_{j_1}\dots\del\phi_{j_n}} = \sum_{k<l} \frac{d^n}{dx^n}\log\left(1+e^x\right)\bigg|_{x={-2\al+2i(\phi_k+\phi_l) \sqrt{B/N}} }\prod_{m=1}^n\left[2i\sqrt{\frac{B}{N}}(\delta_{kj_m} + \delta_{lj_m})\right].
\end{equation}
Notice the Kronecker symbols involving only the two indices $k$, $l$ and one external index $j_m$, so that the derivative tensors vanish unless there are at most 2 distinct values present among the indices $j_m$. The first few orders evaluated at the saddle point are
\allowdisplaybreaks
\begin{align*}
\frac{\del^2 S}{\del\phi_k\del\phi_l}&= 
	\begin{cases}
	    -\frac{4B}{N} \phi_0 (1-\phi_0) & \text{for } i \neq j,\\
	    -4B\frac{(N-1)}{N} \phi_0 (1-\phi_0) & \text{for } i = j;
	    \end{cases}\\
\frac{\del^3 S}{\del\phi_k\del\phi_l\del\phi_m}&= 
	\begin{cases}
	    0 \hspace{0.5cm} \text{with more than 2 distinct values in $\{k,l,m\}$} ,\\
	    8i\;\left(\frac{B}{N}\right)^{3/2} \phi_0 (1-\phi_0)(1-2\phi_0) \hspace{0.5cm} \text{with 2 distinct values in $\{k,l,m\}$},\\
	    8i\;\left(\frac{B}{N}\right)^{3/2}(N-1)\phi_0 (1-\phi_0)(1-2\phi_0) \hspace{0.5cm} \text{all indices are the same;}
	    \end{cases}\\
\frac{\del^4 S}{\del\phi_k\del\phi_l\del\phi_m\del\phi_n}&=
	\begin{cases}
	    0 \hspace{0.5cm} \text{with more than 2 distinct values in $\{k,l,m,n\}$,}  \\
	    16\; \frac{B^2}{N^2} \phi_0(1-\phi_0)(1-6\phi_0+6\phi_0^2)  \hspace{0.3cm} \text{with 2 distinct values in $\{k,l,m,n\}$,} \\
	    16\; \frac{B^2}{N^2} (N-1) \phi_0(1-\phi_0)(1-6\phi_0+6\phi_0^2) \hspace{0.3cm} \text{all indices are the same.}
	    \end{cases}
\end{align*}
One can in principle build the standard Feynman perturbation theory from this expansion, with the inverse of the matrix ${\del^2 S}/{\del\phi_k\del\phi_l}$ used as the propagator, and the higher-order terms in the $\ph$-expansion used as vertices. However, the expressions for the diagrams will involve summations over indices taking values from $1$ to $N$, contributing positive powers of $N$ that will compete with the negative powers seen in the denominators of the derivative formulas above. Furthermore, the powers of $N$, both in the propagator and the vertices, depend on the number of coincident indices. This leads to complicated accounting of the powers of $N$. By contrast, with the variables $x_J$ and the representation (\ref{Zx}-\ref{Seff}) used in the main text, $N$ is a purely numerical parameter in the action, and the powers of $N$ are recovered straightforwardly --- essentially, by means of Taylor expansions.

\section{Some useful formulas}\label{appsigma}

While the expressions below can be straightforwardly derived from (\ref{sigdef}), they are used extensively in our manipulations and we summarize them for the convenience of the reader:
\begin{align}
&\s_0 = \sum_{J=1}^{\infty} c_J^2 J = -\log(1-\phi_0),\\
&\s_1 = \sum_{J=1}^{\infty} c_J^2 J^2= \frac{1}{1+e^{2\al+4B\phi_0}} = \phi_0 ,\\
&\s_2 = \sum_{J=1}^{\infty} c_J^2 J^3 = \phi_0(1-\phi_0),\\
&\s_3 = \sum_{J=1}^{\infty} c_J^2 J^4  = \phi_0(1-\phi_0)(1-2\phi_0),  \\
&\s_4 = \sum_{J=1}^{\infty} c_J^2 J^5 = \phi_0(1-\phi_0)(1-6\phi_0+6\phi_0^2).
\end{align}

\section{Particle-hole duality}\label{appdual}

The partition function of the two-star model is defined as
\beq
Z\equiv\sum_A e^{-H[A]},\qquad H[A]\equiv \al\sum_{kl}{A_{kl}}+\be\sum_{klm} A_{kl}A_{lm},
\eeq 
We want to change the summation variable as $A_{jk} = 1- \tilde{A}_{jk} -\delta_{jk}$, which precisely corresponds to flipping the state of all the edges (filled to empty, empty to filled). Then,
$$
H[A(\tilde{A})] = \left[-\al - 2\beta(N-1)\right] \sum_{kl}{\tilde{A}_{kl}}+\be\sum_{klm} \tilde{A}_{kl}\tilde{A}_{lm} +
\al N(N-1) + \beta(N^3 - 2N^2 + N).
$$
This means that under this transformation, in the dense regime (substitute $\beta = B/N$), the free energy $F\equiv -\log Z$ changes as
\begin{equation}
F(\al,B) = \al N(N-1) + B(N-1)^2 +F\left(-\al -2B +\frac{2B}{N},B\right).
\end{equation}
Since we express our results as functions of $\phi_0(\al,B)$ defined by (\ref{phiexp}), rather than as functions of $\al$ and $B$, it is handy to correspondingly recast the transformation $\al \rightarrow \tilde{\al}\equiv -\al -2B +\frac{2B}{N}$ as $\phi_0 \rightarrow \tilde{\phi_0}$. To do so, we first express $\al$ through $\phi_0$ as
\begin{equation}
\al  = \frac{1}{2}\log\frac{1-\phi_0}{\phi_0}-2B\phi_0,
\end{equation}
enact the transformation $\al \rightarrow \tilde{\al}$, and then write the equation defining $\tilde\phi_0$ through $\tilde\al$:
\begin{equation}
\tilde\phi_0 = \frac{1}{1+e^{2\tilde\al + 4B\tilde\phi_0}}
= \frac{1}{1+ \frac{\phi_0}{1-\phi_0}e^{-4B(1-\phi_0)+4B/N}e^{4B\tilde{\phi_0}}}.
\end{equation}
This equation can be solved in terms of an $1/N$ expansion as
\begin{equation}
 \tilde\phi_0 \equiv 1-\phi_0 - \frac{1}{N} \frac{4B\phi_0(1-\phi_0)}{1+4B\phi_0(1-\phi_0)} - \frac{1}{N^2}\frac{8\phi_0B^2(1-\phi_0)(1-2\phi_0)}{[1+4B\phi_0(1-\phi_0)]^3}+\cdots.
\label{phi0_trans}
\end{equation}
With this transformation, we write
\begin{equation}
F(\phi_0,B) = N(N-1)\left[\frac{1}{2}\log\frac{1-\phi_0}{\phi_0}-2B\phi_0\right] + B(N-1)^2 + F\left(\tilde\phi_0(\phi_0,B), B\right).
\end{equation}
This relation is respected by our result (\ref{Forder1}) up to order $O(1)$, providing a validation of our derivations.

\section{Monte Carlo sampling}\label{appMC}

We sample graphs from the two-star ensemble using a particular version of Metropolis Monte Carlo algorithm that we have previously employed in \cite{birth} for related models . We summarize it below. General discussion can be found in \cite{MC_Newman}.

We start with an adjacency matrix of an Erd\H{o}s-R\'enyi graph generated by randomly connecting any two vertices with probability $1/2$. Then, for each Monte Carlo step, we propose an update move which flips the current edge (filled to empty, empty to filled) for a randomly picked pair of vertices $(i,j)$. The change in the adjacency matrix can be formally written as 
\begin{equation}
\label{eq:varA}
\Delta_{ij}A_{mn} = (\delta_{im}\delta_{jn}+\delta_{in}\delta_{jm})(1-2A_{ij}).
\end{equation}
This move is then accepted with probability $\text{min}\left(1,e^{H(A) - H(A+\Delta_{ij}A)}\right)$ as per the usual Metropolis algorithm. 

Computing the full Hamiltonian takes a number of operations of order $O(N^2)$, which can be time-consuming when one has to take a large number of samples from a big system. One can optimize the numerical process by computing only the change of the Hamiltonian from each proposed update move. For this, one needs to compute the change in the sum of degrees $\sum_{k}d_k = \sum_{kl}A_{kl}$, and in $\sum_{k}d_k^2 = \sum_{klm}A_{kl}A_{lm}$. From \eqref{eq:varA}, the change in degree of node $m$ is
\begin{equation}
\Delta_{ij} d_m = (\delta_{im}+\delta_{jm})(1-2A_{ij}).
\end{equation}
It follows that
\beq
\Delta_{ij}\sum_m d_m = 2(1-2A_{ij}),\qquad
\Delta_{ij}\sum_{m}d^2_{m} = 2\left[(1-2A_{ij})(d_i + d_j)
+1\right].
\eeq
The change in the Hamiltonian is thus
\begin{equation}
\Delta_{ij}H(A)= H(A+\Delta_{ij}A)-H(A) = -2(1-2A_{ij})(\al + \beta(d_i+d_j)+\beta).
\end{equation}
This allows one to compute the acceptance probability with a number of operations of order 1.

We compute the mean degree $\left<k\right>$ and the mean degree squared $\left<k^2\right>$ by averaging over $5\times10^5$ Monte Carlo sample points, which are taken at intervals of $N^2/10$ elementary Monte Carlo steps with the update rule described above. The long interval between the samples ensures that a significant part of the graph gets updated by the stochastic evolution before each next sample is taken.


\end{document}